\documentstyle[12pt]{article}
\newcounter{abc}
\newcommand{\gE}{\mbox{$\tau$}}
\newcommand{\gM}{\mbox{$h$}}
\newcommand{\oE}{\mbox{${\cal E}$}}

\newcommand{\oM}{\mbox{${\cal M }$}}
\newcommand{\oMc}{\mbox{${\cal M }_c$}}
\newcommand{\aE}{\mbox{$a_{{\cal E}}$}}
\newcommand{\aM}{\mbox{$a_{{\cal M}}$}}
\newcommand{\lM}{\mbox{$\lambda_{{\cal M}}$}}
\newcommand{\lE}{\mbox{$\lambda_{{\cal E}}$}}
\newcommand{\pME}{\mbox{$p_{{\cal M},{\cal E}}$}}

\newcommand{\pMEME}{\mbox{$p_{{\cal M},{\cal E}}({\cal M},{\cal E})$}}

\newcommand{\ptM}{\mbox{$\tilde{p}_{{\cal M}}$}}
\newcommand{\ptMstar}{\mbox{$\tilde{p}_{{\cal M}}^{\star}$}}

\newcommand{\pMM}{\mbox{$p_{{\cal M}}({\cal M})$}}

\newcommand{\mix}{\mbox{$s$}}
\newcommand{\mixp}{\mbox{$r$} }
\newcommand{\dfac}{\mbox{$\frac{1}{1-\mix \mixp}$}}
\newcommand{\pr}{\mbox{$p_L(\rho )$}}
\newcommand{\er}{\mbox{$u(\rho )$}}
\newcommand{\uc}{\mbox{$u_c$}}

\begin{document}
\title{Critical point field mixing in an asymmetric\\ lattice gas model}
\author{N. B. Wilding \\
\small{Institut f\"{u}r Theoretische Physik, Philosophenweg 19,}\\ {\small
Universit\"{a}t
Heidelberg, D-6900 Heidelberg, Germany}}
\date{}
\setcounter{page}{0}
\maketitle

\begin{abstract}

The field mixing that manifests broken particle-hole symmetry, is studied
for a 2-D asymmetric lattice gas model having tunable field mixing
properties.  Monte Carlo simulations within the grand canonical ensemble are
used to obtain the critical density distribution for different degrees of
particle-hole asymmetry.  Except in the special case when this asymmetry
vanishes, the density distributions exhibit an antisymmetric correction to
the limiting scale-invariant form.  The presence of this correction reflects
the mixing of the critical energy density into the ordering operator.  Its
functional form is found to be in excellent agreement with that predicted by
the mixed-field finite-size-scaling theory of Bruce and Wilding.  A
computational procedure for measuring the significant field mixing parameter
is also described, and its accuracy gauged by comparing the results with
exact values obtained analytically.

\end{abstract}

\thispagestyle{empty}
\begin{center}
PACS numbers 64.70, 64.70F
\end{center}
\newpage

\section{Introduction}

It has long been appreciated \cite{REHR1} that the lack of symmetry between
the liquid and vapour phases of a fluid leads to a mixing of the temperature
and chemical potential in the two relevant scaling fields close to the
critical point.  This is in marked contrast to the situation for models of
the Ising symmetry such as the ordinary lattice gas \cite{LEE}, which as a
consequence of their so-called `particle-hole' symmetry\footnote{A lattice
gas possesses particle-hole symmetry if the Hamiltonian, a function of the
site occupation numbers $\sigma_i=0,1$ satisfies the relation ${\cal H}
\rightarrow {\cal H}+A\sum_i\sigma_i+B$ (with A, B constants) under the
transformation $\sigma_i \rightarrow 1-\sigma_i$.  Real (continuous) fluids
lack particle-hole symmetry {\it ipso-facto} since a hole is not defined.},
exhibit no field mixing.  Although the reduced symmetry of fluids has no
consequences for their universal properties (which for systems with
short-ranged interactions correspond to the Ising universality class), it is
predicted that certain {\em non-universal} effects are engendered by field
mixing.  Principal among these, is the existence of a weak energy-like
singularity of the coexistence diameter on the approach to criticality.  The
presence of this singularity, now firmly established experimentally
\cite{SENGERS,JUNGST}, constitutes a failure for the century-old empirical
`law of rectilinear diameter'. Indeed, some theoretical progress has been made
towards an understanding of the microscopic factors governing the amplitude
of the diameter singularity \cite{GO1,GO2}.

Recently however, a new finite-size-scaling theory has been developed that
relates the mixed character of the fluid scaling fields to the interplay of
the near-critical energy and density fluctuations \cite{BRUCE1,WILDING1}.
The theory predicts that as a result of field mixing, the energy operator
features in the critical density distribution, giving rise (as outlined
below) to a correction to the limiting (large $L$) universal form of the
density distribution.  This correction is subdominant to the limiting form
and dies away with increasing $L$.  To leading order in the theory, its
functional form is prescribed by independently known functions
characteristic of the Ising universality class.  Moreover the symmetry of
the correction (an antisymmetric function) differs from that of the limiting
form (a symmetric function).  Its presence in the density distribution is
therefore a potentially distinctive signature of the field mixing
phenomenon, one that can in principle be isolated and analysed by means of
computer simulation measurements of the density fluctuations.  Indeed, the
potential utility of the theory in facilitating simulation studies of field
mixing was clearly demonstrated in an extensive Monte Carlo investigation of
the 2-D Lennard-Jones fluid near the liquid-vapour critical
point\cite{WILDING1}.  Measurements of the near-critical density
distribution yielded an antisymmetric field mixing component that mapped
quite well onto the predicted universal form.  In addition, a computational
prescription was set out for estimating the more significant of the two
field mixing parameters (that which controls the extent to which the
chemical potential features in the thermal scaling field).

Notwithstanding the successes of the study reported in \cite{BRUCE1,
WILDING1}, one is still confronted with a number of difficulties if one
wishes to assess the quantitative validity of the new theory.  The most
severe problem is the computational burden imposed by simulations of
realistic fluid models (such as the Lennard-Jones system), which owing to
their continuous, long-ranged interaction potential, require intensive
floating-point calculations.  The computational demands of such simulations
far exceed those of lattice-based particle or spin systems.  Moreover, since
exact values of the critical couplings are not generally available for
realistic fluid models, it is difficult to probe the asymptotic critical
region.  This can complicate the isolation of the field mixing correction to
the density distribution, which must be identified amidst other corrections
associated with small departures from criticality.  The analytical
intractability of realistic fluid models also precludes an assessment of the
accuracy of any computational procedure for measuring the significant mixing
parameter.

Clearly, in order to facilitate a more detailed assessment of the
mixed-field finite-size-scaling theory, it would be beneficial to work with
a model system having fewer of the drawbacks listed above.  One such system
is the 2-D asymmetric lattice gas model, originally proposed by Mermin, which
whilst being comparatively much easier to tackle computationally than e.g
the Lennard-Jones fluid, is also analytically solvable for the critical
point couplings and field mixing parameters.  It therefore provides an ideal
test-bed for the theory, permitting an accurate determination of the field
mixing correction, and providing a benchmark against which, the accuracy of
any computational method for determining the field mixing parameters can be
gauged.

Below we describe the results of field mixing studies of Mermin's asymmetric
lattice gas model.  The layout of the paper is as follows: the model is
described in section \ref{sec:back} and set within the framework of the
mixed-field finite-size-scaling theory.  Exact values for the significant
mixing parameter of the model are also calculated.  In section
\ref{sec:res}, Monte Carlo measurements of the critical density distribution
are presented.  The antisymmetric component of these distributions is
isolated and compared with the predicted form of the field mixing
correction.  Estimates of the significant mixing parameter are also deduced
from the simulation data and compared with the exact results.  Section
\ref{sec:concs} details our conclusions.

\section{Background}
\label{sec:back}

The asymmetric lattice gas model forming the focus of the present work, was
first proposed by Mermin \cite{MERMIN1} as an example of a system exhibiting
a singular coexistence diameter\footnote{We note in passing, that the model
also possesses a rich phase structure that has been investigated in detail
by other workers \protect \cite{REHR2,ZOLLWEG1,MULHOLLAND1,MULHOLLAND2}}.
The model consists of an ordinary 2-D square lattice gas (whose nearest
neighbour coupling we denote $J$) in which particle-hole symmetry is
manually destroyed by forbidding occupation of sites whose row and column
numbers are both even.  By this action one creates two types of sites:
odd-odd sites having coordination number 4, and odd-even (or even-odd) sites
having coordination number 2.  It is straightforward to show that the
particle-hole symmetry that obtains in the ordinary lattice gas is
equivalent to the requirement that all sites have the same average energy
environment.  The presence of two sets of inequivalent sublattices in the
asymmetric lattice gas, clearly violates this condition and leads to field
mixing.  It transpires, however, that if one introduces an additional
coupling $K$, between atoms on the sublattice of odd-odd sites, then the
degree of particle-hole asymmetry can be {\em tuned}.  Indeed for the
special choice $K=-J/2$, the average energy per site becomes equal for both
sublattices and consequently particle-hole symmetry is once more restored.

Aside from its field mixing properties, the chief asset of Mermin's
asymmetric lattice gas, is its analytic tractability.  The grand partition
function of the asymmetric model can be related by means of analytic
transforms to that of the ordinary lattice gas, for which in turn a wealth
of exact results are known in two dimensions \cite{LEE}.  Specifically, one
finds
\cite{MERMIN1}:

\begin{equation}
\Omega (\mu,T)=(1+e^{\mu/kT})^{2{\overline N}}{\overline \Omega} ({\overline
\mu},{\overline T})
\label{eq:part}
\end{equation}
where $\Omega$ is the partition function of the asymmetric model and $\mu$
and $T$ are the chemical potential and temperature respectively.  Bars
denotes quantities in the ordinary lattice gas and ${\overline N}=N/3$ where
$N$ is the number of allowed sites in the asymmetric model.

Introducing the dimensionless chemical potential $\xi=\mu/k_bT$ and
dimensionless coupling parameters $\eta=J/k_bT$, $\lambda=K/k_bT$,
equation~\ref{eq:part} leads to the following relationships \cite{MERMIN1}:

\setcounter{abc}{1}
\begin{eqnarray}
{\overline \xi} & = & \xi + 4\ln\left [ \frac{1+e^{\xi+
\eta}}{1+e^\xi}\right]\\
\addtocounter{equation}{-1}
\label{eq:relnsa}
\addtocounter{abc}{1}
{\overline \eta} & = & \lambda + \ln \left [\frac{(1+e^\xi
)(1+e^{\xi+2\eta})}{(1+e^{\xi+\eta })^2}\right ]
\label{eq:relnsb}
\end{eqnarray}
\setcounter{abc}{0}
where ${\overline \xi}={\overline \mu}/k_b{\overline T}$ and ${\overline
\eta}={\overline J}/k_b{\overline T}$ are respectively the dimensionless
chemical potential and dimensionless nearest neighbour coupling constant of
the ordinary lattice gas.

Now, it transpires \cite{LEE} that in the ordinary lattice gas the
liquid-vapour coexistence line is specified by ${\overline \xi}=-2{\overline
\eta}$, while the critical point that terminates this line is given by the
solutions to the relations:

\begin{equation}
\sinh({\overline \eta_c}/2)=1, \hspace{1cm} {\overline \xi_c}=-2{\overline
\eta_c}
\label{eq:critpars}
\end{equation}
Setting ${\overline \xi}=-2{\overline \eta}$ in equations~\ref{eq:relnsa}
and \ref{eq:relnsb} then yields the coexistence condition for the asymmetric
model:

\begin{equation}
\xi + 2\lambda+2\ln\left [ \frac{1+e^{\xi + 2\eta}}{1+e^\xi}\right ] = 0
\label{eq:coexsur}
\end{equation}
which represents a {\em surface} in the space of $\xi$, $\eta$ and
$\lambda$.  Note however, that since $\lambda$ and $\eta$ both enter only as
multiplicative factors in the configurational energy (see
equation~\ref{eq:pot} below), this coexistence surface can be represented in
terms of a family of coexistence {\em curves} in the space\footnote{Note
that we might equally well have chosen $\lambda$ instead of $\eta$ as the
independent coupling variable.} of $\xi$ and $\eta$, each curve being
parameterised by a different value of the coupling ratio $\lambda/\eta$.
This ratio is consequently the crucial parameter controlling the coexistence
and field mixing properties.

We shall also find it useful to obtain the critical density of the
asymmetric lattice gas, which can be calculated by appeal to the relation:

\begin{equation}
\rho_c=\frac{1}{N}\left. \frac{\partial \ln \Omega (\mu,T)}{\partial \mu}
\right |_c
\end{equation}
from whence, one obtains \cite{MERMIN1}:

\begin{equation}
\rho_c = \frac{2}{3}\left [ 1+e^{-\xi} \right ]^{-1} + \frac{1}{3}{\overline
\rho_c} \left ( \frac{\partial {\overline \xi}}{\partial \xi} \right )_c +
\frac{1}{3}{\overline u_c} \left (\frac{\partial {\overline \eta}}{\partial
\xi}
\right )_c
\label{eq:rhoc}
\end{equation}
where the subscript c on the derivatives signifies that they are to be
evaluated at criticality.  The quantities ${\overline u_c}$ and ${\overline
\rho_c}$ are respectively the critical energy density and number
density of the ordinary lattice gas model, the particle-hole symmetry of
which implies ${\overline \rho_c}=0.5$.  The value of
${\overline u_c}$ is also known exactly; it is related to the critical
energy density $u^I_c$ of the 2-D Ising model by the relation ${\overline
u_c}=\frac{1}{4}[2+u^I_c]$, where from Onsager's solution,
$u^I_c=\sqrt{2}/2$.  The derivatives in equation~\ref{eq:rhoc} can be
calculated straightforwardly to yield

\setcounter{abc}{1}
\begin{eqnarray}
\frac{\partial {\overline \xi}}{\partial \xi} & = & \frac{1
+ 5 e^{\xi + \eta} - 3 e^{\xi} + e^{2\xi + \eta}}{ (1 + e^{\xi}) (1 + e^{\xi +
\eta})} \\
\addtocounter{equation}{-1}
\addtocounter{abc}{1}
\frac{\partial {\overline \eta}}{\partial \xi} & = &
\frac{e^{\xi} - e^{2 \xi + \eta} + 2 e^{2 \xi + 2 \eta}+ e^{\xi + 2 \eta} -
e^{2 \xi + 3
\eta} - 2 e^{\xi + \eta}}{(1 + e^{\xi}) (1 +e^{\xi + 2 \eta}) (1 + e^{\xi +
\eta})}
\label {eq:derivs}
\end{eqnarray}
\setcounter{abc}{0}

Substituting for ${\overline \eta_c}$ and ${\overline \xi_c}$ in
equations~\ref{eq:relnsa} and \ref{eq:relnsb}, and feeding the results for
$\eta_c$ and $\xi_c$ into equation~\ref{eq:rhoc}, one readily obtains the
critical density $\rho_c$ as a function of $\lambda/\eta$.  This
relationship is shown in figure \ref{fig:densgraph} for values of
$\lambda/\eta$ in the range $(-1/2,1/2)$, which, as will be seen,
encompasses a wide range of field mixing behaviour.  The figure
clearly demonstrates that when $\lambda/\eta=-1/2$ the critical density is
that of the ordinary lattice gas (${\overline \rho_c}=0.5$), thus confirming
that particle hole symmetry is restored for this value of the coupling
ratio.  It is also evident that increasing $\lambda/\eta$ causes the
critical density to decrease monotonically.  The nature of this density
shift finds illustration in the simulation results to be described later.

We turn now to the field mixing properties of the asymmetric lattice gas
model, which we analyse within the framework of the mixed-field
finite-size-scaling theory of references \cite{BRUCE1,WILDING1}.  To this
end, we consider a 2-D system of side $L$, having a maximum available volume
$V=3L^2/4$.  The configurational energy $\Phi(\{\sigma\})$ is given by

\begin{equation}
\Phi(\{\sigma\}) = \sum_{<i,j>}\eta\sigma_i\sigma_j +
\sum_{[m,n]}\lambda\sigma_m\sigma_n
\label{eq:pot}
\end{equation}
with $\sigma_i=0,1$.  The site indices $i$ and $j$ are taken to run over all
allowed nearest neighbour sites, while $m$ and $n$ run only over nearest
neighbours on the sublattice of odd-odd sites.  We assume also that the system
is
thermodynamically open so that the particle number density
$\rho=\sum_i\sigma_i/V$ can fluctuate.  In this paper, we shall be concerned
with the statistical behaviour of both the number density $\rho$, and the
configurational energy density $u=V^{-1}\eta^{-1}\Phi(\{\sigma\})$, the
latter of which we write in units of the coupling parameter $\eta$.

For a given choice of the coupling ratio $\lambda/\eta$, the critical point
is located by critical values of the reduced chemical potential $\xi_c$ and
reduced coupling $\eta_c$.  Deviations of $\xi$ and $\eta$ from their
critical values determine the size of the two relevant scaling fields
\cite{WEGNER}.  In the absence of particle-hole symmetry, it is expected
\cite{REHR1} that the relevant scaling fields comprise (asymptotically)
linear combinations of these deviations:

\begin{equation}
\tau = \eta_c-\eta +s(\xi - \xi_c) \hspace{1cm} h=\xi - \xi_c+r(\eta_c-\eta )
\label{eq:scaflds}
\end{equation}
where $\tau$ is the temperature-like scaling field and $h$ is the field-like
scaling field.  The parameters \mix\ and \mixp\ are system-specific
quantities controlling the degree of field mixing.  In particular, \mixp is
identifiable as the limiting critical slope of the coexistence curve in the
space of $\xi$ and $\eta$ \cite{REHR1}.  The role of the parameter \mix\
(which controls the degree to which the chemical potential features in the
thermal scaling field) is, however, more significant: it determines the size
of the diameter singularity.  We term \mix\ the significant mixing
parameter.

Conjugate to the two relevant scaling fields are scaling operators ${\cal
E}$ and ${ \cal M}$, which are found to comprise linear combinations of
the energy and number densities \cite{WILDING1}:

\begin{equation}
\oE  =  \dfac \left[  u  - r \rho \right] \hspace{1cm}\oM  = \dfac \left[  \rho
- s u \right]
\label{eq:oplinks}
\end{equation}
where \oE\ is the energy-like operator and \oM\ the ordering operator.  In
the Ising context (for which $\mix = \mixp =0$), \oM\ is simply the
magnetisation while \oE\ is the energy density.

Near criticality, the joint probability distribution \pMEME\ of the
operators \oE\ and \oM\ is expected to exhibit scaling behaviour.  In
particular, in the limit of large system size $L$, the distribution of the
ordering operator $\pMM = \int \pMEME d\oE$ is expected to be describable by
a finite-size-scaling relation having the form \cite{BINDER}:

\begin{equation}
\pMM \simeq \aM ^{-1}L^{d -\lambda _{{\cal M}} } \tilde{p}_{\cal M} (\aM
^{-1}L^{d -\lambda _{{\cal M}} } \delta \oM , \aM L^{\lambda_{\cal M} }\gM ,\aE
L^{\lambda _{{\cal E}} }\gE) \label{eq:ansatz}
\end{equation}
where $\delta \oM \equiv \oM - \oMc$ and the function \ptM\ is predicted to
be {\em universal}, modulo the choice for the scale-factors \aM\ and \aE\
of the two relevant fields, whose scaling indices are $\lM = d-\beta/\nu$
and $\lE=1/\nu$ respectively.  This scaling form (which has its basis in the
renormalisation group scaling properties of the multi-point correlation
functions
\cite{BRUCE2}) is well-supported by Monte Carlo studies of 2-D
Ising and $\phi ^4 $ models, where the ordering operator \oM\ is simply the
magnetisation \cite{BRUCE3}.

In asymmetric systems, the mixing of the energy density into \oM\ precludes
direct simulation measurements of \pMM , since in general the value of \mix\
will not be known {\it a-priori}.  Instead, it is expedient to focus on the
distribution of the {\em density}, which is obtained from the
joint distribution of the mixed operators by integrating over the energy
spectrum and expanding in the mixing parameter \mix\ :

\begin{equation}
\pr  =  \int \pME ( \rho - \mix \oE , \oE ) d\oE
\end{equation}
from which one finds \cite{WILDING1}

\begin{equation}
\pr \simeq \aM ^{-1}L^{d -\lambda _{{\cal M}} } \ptM (\aM ^{-1}L^{d -\lambda
_{{\cal M}} } [\rho - \rho _c ] ,
\aM L^{\lambda_{\cal M}} \gM , \aE L^{\lambda_{\cal E}} \gE ) +\Delta \pr
\label{eq:dendst}
\end{equation}
with
\begin{equation}
\Delta \pr = - \mix \frac{\partial }{\partial \rho}\left \{\pr \left [<\er > -
\uc  - \mixp (\rho -\rho _c ) \right ] \right \} +O(\mix
^2) \nonumber
\label{eq:dencor}
\end{equation}
where $<\er >$ (hereafter referred to as the energy function) is the mean
energy density for a given $\rho$.

The function $\Delta \pr$ describes (to linear order in \mix ) the component
of the critical density distribution associated with field mixing.
Precisely at criticality, equation~\ref{eq:dendst} may be written in the
form:

\begin{equation}
\pr \simeq \aM ^{-1}L^{\beta/\nu } \left[\ptMstar(x) - s\aE
\aM ^{-1} L^{-(1-\alpha-\beta)/\nu}\frac{\partial}{\partial x}
\left( \ptMstar (x) \tilde{\epsilon}^{\star}(x)\right) +
O(s^2) \right]_{x=\aM ^{-1}L^{\beta/\nu } [\rho - \rho _c ] }
\label{eq:prediction}
\end{equation}
where $\ptMstar (x) \equiv \ptM(x,0,0)$ and

\begin{equation}
\tilde{\epsilon }^{\star }(x) \equiv \left. \frac{\partial \ln \ptM
(x,0,z)}{\partial
z} \right |_{z=0}  =  \aE^{-1}L^{d-1/\nu}[<\er > - \uc  -  \mixp (\rho -\rho _c
)] + O(s)
\label{eq:er3}
\end{equation}
is a universal function whose form has (like that of \ptMstar ) been
previously established in Monte Carlo studies of the critical 2-D Ising
model \cite{BRUCE3}, where it is simply the energy function for the
magnetisation.

To leading order in \mix , the critical density distribution can thus be
expressed as a sum of two independently-known universal components.  The
first of these, $\ptMstar (x)$, is a function having the same form as the
critical magnetisation distribution of the Ising model.  The second,
$\frac{\partial}{\partial x} \left( \ptMstar (x)
\tilde{\epsilon}^{\star}(x)\right)$, is a function characterising the
critical energy operator and represents (to linear order in \mix ) the field
mixing contribution to the density distribution.  This field mixing term is
down on the first term by a factor $L^{-(1-\alpha -\beta)/\nu}$ and
therefore represents a {\em correction} to the large $L$ limiting behaviour.
Given further the symmetries of $\tilde{\epsilon }(x)$ and $\ptMstar (x)$,
both of which are even (symmetric) in the scaling variable $x$, the field
mixing correction is the leading {\em antisymmetric} contribution to the
density distribution.  Accordingly, it can be isolated from measurements of
the critical density distribution, simply by antisymmeterising around
$\rho_c$.

In addition to furnishing the functional form of the field mixing
correction, equation~\ref{eq:dencor} also constitutes a computational
prescription for estimating the significant mixing parameter \mix\ .  The
value of \mix\ is simply the single scale factor required to match the
critical point form of the measured function $\Delta \pr=-\mix\frac{\partial
}{\partial \rho}\left \{ \pr \left [<\er > - \uc - \mixp (\rho -\rho _c )
\right ] \right \}$ to the measured antisymmetric component of the critical
density distribution.  In the present case, the accuracy of this procedure
can be gauged, since the exact value of \mix\ is obtainable analytically.

An exact calculation of \mix\ follows from the observation that in the
ordinary lattice gas, the field-like scaling field \gM\ coincides with the
line ${\overline \eta}={\overline \eta_c}$ in the space of ${\overline \xi
}$ and ${\overline \eta}$.  It follows that in the asymmetric model, the
direction of \gM\ in the space of $\xi$ and $\eta$ can be obtained from
equation~\ref{eq:relnsb} by setting ${\overline \eta}={\overline \eta_c}$,
$\lambda=\lambda(\eta)$ and solving for $\eta$.  The value of \mix\ is then
given by

\begin{equation}
\mix =\left ( \frac{\partial \eta }{\partial \xi}\right )_c
\end{equation}
where the derivative is to be evaluated at criticality.  The results of this
calculation are displayed in figure~\ref{fig:sgraph} as a function of
$\lambda/\eta$ in the range $(-1/2,1/2)$.  As anticipated, \mix\ vanishes
for $\lambda/\eta =-1/2$.  We note also the presence of a broad minimum in
the value of \mix\ at $\lambda/\eta \simeq -0.08$.

\section{Monte Carlo studies}
\label{sec:res}

\subsection{Computational details}

The Monte Carlo simulations reported here, were all performed using a
Metropolis algorithm within the grand canonical ensemble.  Two system sizes
were studied, having linear dimension $L=20$ and $L=30$.  Periodic boundary
conditions were employed throughout.  The basic observables recorded were
the probability distribution of the number density $P_L(\rho)$ and the
energy function $<u(\rho)>$.  The distribution of the density was obtained
initially as a histogram.  The energy function was accumulated as an average
of the energy for each value of $\rho$ explored in the course of the
simulation.  All simulations were performed at the exact critical point of
the model, obtained (for a given choice of $\lambda/\eta$) from
equations~\ref{eq:relnsa}, \ref{eq:relnsb} and \ref{eq:critpars}.  The
$L=20$ systems consisted of $1\times 10^6$ lattice sweeps for equilibrium,
followed by a sequence of $2\times 10^5$ observations with $100$ sweeps
between each observation.  For the $L=30$ system, equilibration times of
$4\times 10^6$ sweeps were used with $2\times 10^5$ observations separated
by $200$ lattice sweeps.  In each instance the whole procedure was repeated
$12$ times to test the statistical independence of the data and to assign
statistical errors to the results.

\subsection{Results}

Measurements of the critical density distribution were collected for the
$L=20$ system at five distinct values of $\lambda/\eta$, namely $-1/2, -1/4,
0, 1/4, 1/2$.  For the $L=30$ system, measurements were made with
$\lambda/\eta =-5/12$ and $\lambda/\eta=0$.  The $L=20$ distributions are
shown in figure~\ref{fig:dists}, where each has been normalised to unit
integrated weight.  With the exception of the case $\lambda/\eta=-1/2$, the
distributions display a marked asymmetry.  Also apparent, is a pronounced
shifting of the high density peak to successively lower densities as
$\lambda/\eta$ is increased.  By contrast, however, the position of the low
density peak is relatively unaffected by changes in $\lambda/\eta$.
Measurements of the critical density (calculable simply as the first moment
of these distributions), agree well with the exact values obtained from
equation~\ref{eq:rhoc}.

That the asymmetry of the distributions vanishes for $\lambda/\eta=-1/2$, is
demonstrated in figure~\ref{fig:map} where a comparison is shown between the
critical density distribution and the magnetisation distribution of the 2-D
spin-$\frac{1}{2}$ Ising model at its exact critical point, as previously
obtained by Nicolaides and Bruce \cite{BRUCE3}.  The density data has been
expressed in terms of the scaling variable $x=\aM ^{-1}L^{\beta/\nu } [\rho
- \rho _c ]$, and the data collapse has been effected by a single scaling
(choice of the non-universal scale factor $\aM$) such that both distributions
have unit variance.  Clearly, the overall quality of the data collapse is
impressive, except for small departures near the vestiges of the density
distribution.  These however, can be traced to the fact that the $L=20$
system is not quite sufficiently large to allow proper sampling of the
high and low density tails of the density distribution.  For this reason,
detailed analysis of the corrections to the density distributions was
carried out for the $L=30$ system where this problem is much less evident.

As previously noted, the antisymmetric component of the density
distributions may be isolated simply by antisymmeterising the density
distribution about $\rho_c$, whose value is obtainable from
equation~\ref{eq:rhoc}.  The results of applying this procedure are depicted
in figure~\ref{fig:anticomp} for the $L=30$ system at $\lambda/\eta=-5/12$
and $\lambda/\eta=0$.  In the former case, the size of the antisymmetric
component is quite small, while in the latter it is considerably larger.
These antisymmetric components are replotted as functions of the scaling
variable $x$ in figure~\ref{fig:corrmap}, together with the predicted
universal form of the field mixing correction, prescribed in
equation~\ref{eq:prediction} and obtained from Ising model studies
\cite{BRUCE3,WILDING1}.  The data have all been brought into coincidence by
single scalings of the non-universal scale factor \aM\ and the ordinate.
In both instances the antisymmetric component of the density distribution
maps extremely well onto the predicted universal form.

Turning now to measurements of the significant mixing parameter, the value
of \mix\ can be estimated according to the procedure outlined in the
previous section.  We shall describe this procedure for the case
$\lambda/\eta=-5/12$, for which the antisymmetric component of the density
distribution is rather small.  Some additional complications arise when the
antisymmetric component is larger, and these are discussed separately in the
context of the $\lambda/\eta=0$ data.

In order to measure \mix\ , both the energy function $<\er > -u_c$ and the
density distribution \pr\ are required.  The energy function is displayed in
figures~\ref{fig:enfunc} for the case $\lambda/\eta=-5/12$.  Also included
in figure~\ref{fig:enfunc} is the function $\mixp (\rho-\rho_c)$, where
\mixp (cf equation~\ref{eq:scaflds}) is the limiting critical slope of the
coexistence curve in the plane of $\eta$ and $\xi$, whose value (for a
particular choice of $\lambda/\eta$) is calculable from
equation~\ref{eq:coexsur}.  For the present case, $\lambda/\eta=-5/12$, one
finds $\mixp =-1.054092$.  Now, according to equation~\ref{eq:dencor}, \mix\
is simply the single scale factor required to match the function $\Delta \pr
=-\mix\frac{\partial }{\partial \rho}\left \{ \pr \left [<\er > - \uc -
\mixp (\rho -\rho _c ) \right ] \right \}$ to the antisymmetric component of
the density distribution.  To the extent that $O(s)$ correction to the
energy function and the density distribution can be neglected, the function
$\Delta \pr$ will itself be antisymmetric and thus comparison of the scales
of the two functions can be effected directly, notwithstanding the relative
enhancement of statistical uncertainties in $\Delta \pr$ that results from
taking the numerical derivative.  Figure~\ref{fig:ddcorr} shows both $\Delta
\pr$ and the measured antisymmetric component of the density distribution.
The two data sets have been brought into correspondence by a single scaling
(choice of the ordinate) implying a value $\mix =-0.23(2)$.  This estimate
is in quite good agreement with the exact value of $-0.204...$.

Finally, we address briefly the situation where the antisymmetric component
of the density distribution is not small.  Under these circumstances, $O(s)$
terms make a significant contribution to both \pr\ and \er .  These terms
(manifesting themselves as $O(s^2)$ corrections to $\Delta \pr$) prevent a
direct comparison of $\Delta \pr$ with the measured antisymmetric component
since $\Delta \pr$ will no longer itself be antisymmetric in $\rho-\rho_c$.
Notwithstanding this, an estimate of \mix\ can be obtained by the simple
expedient of antisymmeterising $\Delta \pr$ about $\rho_c$, and comparing
the scales of the two functions.  Carrying out this procedure for the
$\lambda/\eta=0$ data yields an estimate $\mix=-0.38(1)$ to be compared with
the exact value $s=-0.331...$.  The bulk of this discrepancy, some $15\%$,
is presumably attributable to the neglect of field mixing terms of order
$s^2$ and higher, which are not included in equation~\ref{eq:prediction}.

\section{Conclusions}
\label{sec:concs}

The simulation results presented in this paper confirm that in systems with
broken particle-hole symmetry, the field mixing phenomenon gives rise to an
antisymmetric correction to the limiting form of the critical density
distribution.  An accurate determination of the functional form of this
correction has demonstrated it to be in excellent quantitative agreement
with the prediction of the mixed-field finite-size-scaling theory of
references \cite{BRUCE1,WILDING1}.  As such the results constitute
substantial corroboration of that theory.

A method for estimating the significant mixing parameter \mix\ was also
assessed, and found to yield accurate results, provided the overall
magnitude of the field mixing correction is not too large.  In view of the
finding \cite{WILDING1} that the size of the field mixing correction in the
Lennard-Jones fluid is indeed rather small (constituting only approximately
$5\%$ by weight of the density distribution for a system of $400$
particles), the method described should in principle permit quite accurate
evaluation of \mix\ in realistic (off-lattice) fluid models.  Of course (and
as noted in the introduction), the feasibility of such studies is contingent
upon an ability to first locate accurately the critical point for the model
of interest.  However, as was demonstrated in reference \cite{WILDING1},
this can also be achieved via a finite-size-scaling analysis of the density
distribution, {\em provided} one works within an ensemble (such as the grand
canonical ensemble) that affords adequate sampling of density fluctuations.

With these points in mind, it would be of considerable interest to extend
the present programme of simulation studies to an investigation of the
microscopic factors influencing the degree of field mixing in realistic
fluid models.  Recent theories have predicted that the size of the
significant mixing parameter is strongly influenced by three-body forces
\cite{GO1,GO2}.  Such interactions could certainly (albeit at greater
computational expense) be incorporated into off-lattice fluid simulations,
allowing their role in the field mixing process to be assessed.

\subsection*{Acknowledgement}

The author is grateful to A D Bruce for helpful communications and for making
available the results of Ising model studies.

\newpage

\begin{figure}[h]
\vspace*{1.5 in}
\caption{The critical density $\rho_c$ of the asymmetric lattice gas model
expressed as a function of the coupling ratio $\lambda/\eta$}
\label{fig:densgraph}
\end{figure}

\begin{figure}[h]
\vspace*{1.5 in}
\caption{The significant field mixing parameter `s' expressed as a function
of the coupling ratio $\lambda/\eta$}
\label{fig:sgraph}
\end{figure}

\begin{figure}[h]
\vspace*{1.5 in}

\caption{The $L=20$ critical density distribution for various values of the
coupling ratio $\lambda/\eta$.  Statistical uncertainties are smaller than
symbol sizes; the curves are simply guides to the eye.  All distributions
are normalised to unit integrated weight.}

\label{fig:dists}
\end{figure}

\begin{figure}[h]
\vspace*{0.5 in}
\label{fig:map}

\caption{The critical density distribution for the $L=20$ system at
$\lambda/\eta=-1/2$, expressed as a function of the scaling variable
$x=L^{\beta/\nu} a^{-1}_M(\rho-\rho_c)$.  Also shown (full curve) is the
critical magnetisation distribution of the 2-D Ising model obtained in
\protect \cite{BRUCE3}.  The non-universal scale factor implicit in the
definition of the scaling variable, has been chosen so that both
distributions have unit variance.}

\end{figure}

\begin{figure}[h]
\vspace*{0.5 in}

\caption{The measured antisymmetric component of the $L=30$ critical density
distributions at $\lambda/\eta=-5/12$ and $\lambda/\eta=0$.  The lines are
guides to the eye.}

\label{fig:anticomp}
\end{figure}

\begin{figure}[h]
\vspace*{0.5 in}

\caption{The data of figure~\protect\ref{fig:anticomp} re-expressed in terms
of the scaling variable $x=L^{\beta/\nu} a^{-1}_M(\rho-\rho_c)$ and shown as
the data points.  The solid curve represents the prediction following from
equation \protect\ref{eq:prediction}, utilising predetermined Ising forms
\protect\cite{BRUCE3}.  The measured correction data have been brought into
coincidence with the predicted form via appropriate choices of the
non-universal scale factor \aM\ and the ordinate.}

\label{fig:corrmap}
\end{figure}

\begin{figure}[h]
\vspace*{0.5 in}

\caption{The measured form of the critical energy function $<u(\rho\ )>-u_c$
for the $L=30$ system at $\lambda/\eta=-5/12$.  Statistical uncertainties do
not exceed the symbol sizes.  Also shown (solid line) is the function
$\mixp(\rho-\rho_c)$, with $r=-1.054092$.}

\label{fig:enfunc}
\end{figure}

\begin{figure}[h]
\vspace*{0.5 in}

\caption{The measured function $\Delta \pr =-\mix\frac{\partial }{\partial
\rho}\left \{ \pr \left [<\er > - \uc - \mixp (\rho -\rho _c ) \right ]
\right \}$ for the $L=30$ system at $\lambda/\eta=-5/12$ and shown as
crosses $(\times )$.  Also shown as circles ($\circ$) is the measured
antisymmetric component of the critical density distribution, cf.
figure~\protect\ref{fig:anticomp}.  The matching shown was effected by a
choice of the ordinate implying $\mix=-0.23(2)$.}

\label{fig:ddcorr}
\end{figure}

\end{document}